\documentclass{vietnam}

\def\be{\begin{equation}}
\def\ee{\end{equation}}
\def\bea{\begin{eqnarray}}
\def\eea{\end{eqnarray}}

\usepackage{amsmath,amssymb,mathtools,bm}
\usepackage{booktabs}
\usepackage{braket}
\usepackage{multirow}
\newcommand{\lr}[1]{\left(#1\right)}
\newcommand{\tti}[1]{\text{\tiny{#1}}}

\newcommand{\Photo}{}

\begin{document}
\vspace*{4cm}
\title{Gravitational Waves from the Hagedorn Phase}

\author{ Gonzalo Villa }

\address{DAMTP,University of Cambridge, Wilberforce Road,  \\
Cambridge, CB3 0WA, UK}

\maketitle\abstracts{
We illustrate the main points behind the computation of the gravitational wave spectrum originated from a phase in the early universe where the energy density is dominated by highly excited fundamental string degrees of freedom - a Hagedorn phase.
This phase is described using the Boltzmann equation approach to string thermodynamics, which we illustrate via a toy model which permits analytic computations for the evolution of perturbations.
This window into the out-of-equilibrium regime allows us to compute equilibration rates and conclude that, in realistic scenarios, long, open strings dominate the ensemble, source gravitational waves and provide a succesful reheating.
Furthermore, we compare the results against the Standard Model prediction for a gravitational wave background of thermal origin and conclude that the string prediction is larger.
This is one of the few cases in which a signal of stringy origin dominates over the field theory analogue.
This work is based on~\cite{Frey:2023khe,Frey:2024jqy} and is a contribution to the Proceedings of the 29th symposium on Particles, Strings and Cosmology (PASCOS 2024).}

\section{PASCOS: PArticles, Strings, COSmology}

As far as the energies probed at our colliders are concerned, the microscopic behaviour of nature is well described by a gauge quantum field theory, the Standard Model (SM).
As is well known, however, a consistent description of gravity at high energies within the framework of quantum field theory remains elusive.
String theory, on the other hand, maintains consistency at high energies whilst reproducing gravity coupled to quantum field theory at low energies, together with strict consistency constraints that the interactions must satisfy.
This makes the framework attractive and worth studying on itself; however, it is important to extract concrete predictions of string models that can be tested.

Even though the connection of string theory to particle physics is clear, it has proven difficult to extract generic predictions that can be tested in colliders.
The problem has two faces: the (presumably) very high string scale
\footnote{Strictly speaking, the string scale is determined dynamically and is therefore model-dependent. In this work we will treat it as a parameter.}
$M_s$, and the landscape of backgrounds in which strings consistently propagate - which leads to a plethora of consistent gauge theories at low energies.
An interesting point of view is to consider cosmology instead, where high energies are naturally reached and string theory predicts phenomena that are surprising (or cannot be described at all) using low-energy reasoning.
The field of study of these novel features in the early universe receives the name of \textit{string cosmology}~\cite{Cicoli:2023opf,Brandenberger:2023ver}.
Furhermore, as a quantum theory of gravitation, a model-independent feature of string theory is the presence of gravitational waves (GWs), and these are therefore an ideal candidate to test the framework.
In this work, we investigate stringy effects in the early universe and a potential GW signature.

\section{The early universe as a GW factory}\label{sec:SM}
Gravity is weak.
This simple statement lies behind the great experimental challenge of probing the quantum nature of the gravitational field, and is encoded in the immensity of the Planck scale $M_p=1/\sqrt{8\pi G}\simeq 10^{19} \text{GeV}$.
Turning the logic around, however, this suggests that high energy processes in the early universe emit a meaningful amount of GWs.
Furthermore, again because of the weakness of gravity, the universe is always transparent to GWs and as such they reach us today carrying information about the process that sourced them.
This motivates \textit{GW astronomy as a way of testing high-energy physics}.

Cosmological backgrounds of GWs~\cite{Caprini:2018mtu} arrive at earth forming a stochastic background - a superposition of uncorrelated signals.
These are described in terms of the energy density in GWs per logarithmic frequency interval 
\begin{equation}\label{eq:SGWB0} 
h^2 \Omega_{\text{\tiny GW},0} = \frac{h^2}{\rho_c}\frac{d \rho_{\text{\tiny{GW}}}}{d \log f} \, , 
\end{equation} 
where $\rho_c=3H_0^2 M_p^2$ is the critical energy density, $H_0 = 100 \, h \, \text{Km} \, \text{sec}^{-1} \, \text{Mpc}^{-1}$ ($h\sim 0.7$) is the Hubble constant today and $\rho_{GW}$ is the energy density in GWs today.
As any contribution to the energy density of the universe sourced before Big Bang Nucleosynthesis (BBN), the integral of $h^2\Omega_{\tti{GW,0}}$ is constrained by the BBN bound to $h^2 \Omega \lesssim 10^{-6}$.

\begin{figure}[t!]
\centering
    \includegraphics[width=0.6\textwidth]{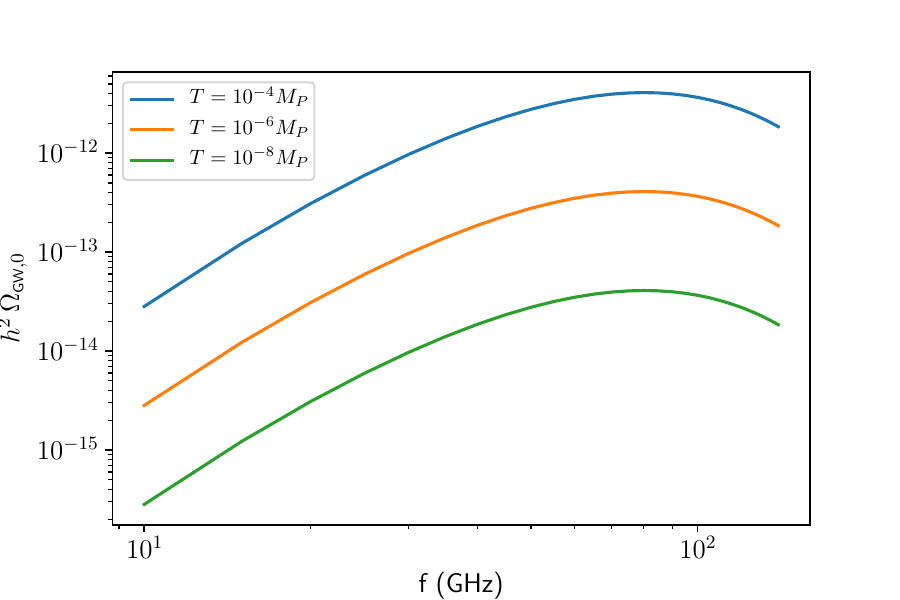}
    \caption{CGMB sourced by the SM for different values of the reheating temperature.}
    \label{fig:SM}
\end{figure}

In this work we will focus on two cosmological sources of GWs at high frequencies (50-100 GHz), noting the tantalizing fact that \textit{the resulting GW spectrum is a UV-sensitive quantity}.
The high-frequency band is currently attracting a lot of attention in both the theory and experimental side~\cite{Aggarwal:2020olq}.
The first such background is the Cosmic Gravitational Microwave
\footnote{Also called Cosmic Gravitational Wave Background in the literature~\cite{Muia:2023wru} since the peak is not neccesarily in the microwave regime, e.g. in nonstandard cosmologies.}
Background (CGMB)~\cite{Ghiglieri:2015nfa,Ghiglieri:2020mhm,Ringwald:2020ist}.
At high-energies, this GW background can be understood as arising from the emision of gravitons by a thermal plasma.
Without delving into details, the resulting contribution to the energy density today per e-fold is of the form
\begin{equation}\label{eq:fraction}
    \frac{d}{d\log a}\left(\frac{d \rho_{\text{\tiny{GW}}}}{d \log f}\right) \sim \frac{T(t)}{M_p} \, \rho(t)\, a(t)^4 \, F(f_{\tti{em}}(t)/T(t))\,.
\end{equation}
where $a(t)$ is the scale factor, $\rho (t)$ is the energy density of the bath, which has temperature $T(t)$ at time $t$.
The function $F(f_{\tti{em}}(t)/T(t))$ is a model-dependent function which generically features a peak when its argument is of order one~\cite{Ringwald:2020ist}, where $f_{\tti{em}(t)}$ is the frequency at the time of emission.
The supression $T/M_p$ penalyses gravitational processes by the ratio of the characteristic scale in the problem, $T$, to the Planck scale.
Two aspects of Eq.~\eqref{eq:fraction} are noteworthy:
\begin{itemize}
\item The peak frequency today is, assuming standard cosmology, the same for every e-fold.
To see this, note that the frequency today only appears in Eq.~\eqref{eq:fraction} in the combination $f_{\tti{em}}/T$ and, neglecting injection of entropy due to degrees of freedom becoming massive, this fraction is constant in standard cosmology.
The CGMB is a therefore a superposition of signals with the same spectral shape at the same frequency.
\item The background releases a fraction of its energy density into GWs, and this fraction is proportional to the temperature $T$.
The only time-dependent quantity in Eq.~\eqref{eq:fraction} is $T$ and thus the shape of the CGMB is dominated by the largest temperatures.
That is, \textit{the shape of the CGMB is an UV-sensitive quantity}, determined by the function $F(x)$ which is itself given by the field content of the high-energy theory.
\end{itemize}
We depict the SM result in Fig.\ref{fig:SM}, for distinct initial temperatures.
We observe that the SM predicts that the shape of the CGMB features a peak around 80 GHz, with an amplitude that depends linearly in the reheating temperature.
Any deviation from such a frequency would thus be an indicator of a deviation from SM physics in the early universe.

The effects of new physics in the shape of the CGMB have been investigated~\cite{Muia:2023wru}.
For the purposes of this work, however, we simply summarize that the qualitative aspects of the SM prediction are very robust to additional degrees of freedom in the early universe.
In addition, at fixed reheating temperature the amplitude predicted by the SM is typically larger than other scenarios with additional high-energy degrees of freedom, a fact that can be understood in terms of entropy injections in the visible sector.
Furthermore, the production of these GWs from a strongly coupled thermal plasma has been studied~\cite{Castells-Tiestos:2022qgu} via holography for the $\mathcal{N}=4$ Super Yang-Mills theory, and the emission rate does not change in order of magnitude as the coupling becomes strong.
We will take this as suggestive evidence that the SM prediction captures the result predicted by a broad class of quantum field theories, including strongly coupled ones.
The SM prediction is therefore robust, (weakly) UV-sensitive and challenging to enhance in field theory.
In the rest of this work we illustrate how string theory provides a counterexample to this conclusion, providing a UV-completion of the CGMB with a larger amplitude.

\section{Boltzmann equations for string thermodynamics}\label{sec:stringthermo}
As the candidate UV-completion of gravity and all other forces, string theory naturally plays a role in the cosmology of the very early universe, where na{\"i}ve extrapolation of the hot big bang should lead to a phase with highly excited strings forming a thermal bath - a \textit{Hagedorn phase}.
Even though this is not neccesarily the case if inflation happened in the early universe and was followed by standard reheating, it has been argued that stringy effects during reheating~\cite{Frey:2005jk} or after brane-antibrane inflation~\cite{Kofman:2005yz} could have given rise to a post-inflationary Hagedorn phase.
It is therefore important to study the thermodynamics of string theory and potential implications in the early universe.

In order to study stringy thermal effects in cosmological settings, we first need to understand the thermodynamics of string theory in Minskowski spacetimes.
The equilibrium description is only valid provided the equilibration rate of the system is much larger than the Hubble scale, which tends to take the system out of equilibrium.
Knowledge of interaction and equilibration rates is thus essential to assess whether equilibrium thermodynamics appropriately describes the gravitating system, and we will adapt the following strategy~\cite{Frey:2023khe}: first, compute the relevant interaction rates at a given order in the string coupling.
Then, use these rates to pose Boltzmann equations and find their equilibrium solutions.
The equilibration rates can then be inferred from the equations describing the dynamics of small perturbations to the equilibrium configurations.
Each step has its own challenges and in this section we will illustrate them in a toy model.

\subsection{The Boltzmann equations for highly excited strings}\label{sec:boltzmann}

In this section, we pose Boltzmann equations for the number distribution of typical
\footnote{The notion of typical string arises from an averaging procedure over all string configurations (states) of a given length~\cite{Frey:2023khe}.}
closed strings, $n(l/l_s)$, where $n_i(x)dx$ is the number of strings in the system with length between $l\equiv x \, l_s$ and $l+dl$, and $l_s=1/M_s$ is the fundamental string length.
In this document we restrict ourselves to the simplest case of very long closed strings propagating on a torus of typical size much smaller than the length of the typical string, although more complicated (and interesting) cases have been studied~\cite{Frey:2023khe}.
Assuming that interactions of long strings should be proportional to their length, and that interactions preserve length, one can write~\cite{Lowe:1994nm}:
\begin{equation}\label{eq:lowe-thorlacius}
\begin{split}
\frac{\partial n(x)}{\partial t}
=&\frac{\kappa}{2}\int_0^x{dy\, \lr{yn(y)(x-y)n(x-y)-xn(x)\vphantom{\frac 12}}}\\ 
 &+\kappa\int_x^\infty{dy\, \lr{yn(y)-(y-x)n(y-x)xn(x)\vphantom{\frac 12}}}\, .
\end{split}
\end{equation}
Where $\kappa l_s$ is a small number controlling the strength of the interactions.
The first line contains interactions with strings shorter than $x$, either a fusion of strings $(y,x-y)\to x$ in the first term, or by self-decay in the second term, where all possible daughter strings have the same probability.
This changes in more general setups, where the string has a probability of decay which can be understood in terms of random walks.
Similarly, the second line contains interactions with strings longer than $x$, either the self decay of strings of length $y$ into $x$ and $y-x$, or the inverse process of fusion of strings $(x,y-x)\to y$.
The assumptions leading to this form of the rates can be shown to be correct by a worldsheet computation in flat space, assuming the strings are much larger than all directions along the torus.
In more complicated cases, the interaction rates admit a geometric interpretation in terms of random walks, on which we will not develop here. 

The equation~\eqref{eq:lowe-thorlacius} is written channel by channel: that is, every line contains a process and its time reversal.
This is a non-linear partial integro-differential equation for which it is generically difficult to find solutions.
In this case, the underlying physics provides simple solutions due to the principle of detailed balance
\footnote{Note that detailed balance is obvious at the microscopic level due to time reversal symmetry of the S-matrix, but not at the coarse-grained level we are studying here.}
: in equilibrium, there is no net transfer of energy among the channels, and so the individual lines must cancel each other.
Writing the equation in this form makes clear how detailed balance is organised, and allows to find equilibrium configurations easily
\footnote{The equilibrium solution in this case can be found~\cite{Lowe:1994nm} by taking a Laplace transform.
This kind of method to solve integral equations is however not available in general.}.
Detailed balance operates length by length, and requires the same condition for both channels,
$n(x)x= n(y)y n(x-y)(x-y)$.
The most general continuous solutions are $n(x)=f(x)/x$ with $f(x)$ unit and exponential functions, but the only physically reasonable ones are those that decay exponentially.
We thus find the equilibrium distribution:
\begin{equation}\label{eq:eq-dist-closed}
    n(x)=\frac{e^{-x/L}}{x}\, .
\end{equation}
The distribution resembles a Boltzmann distribution $n(E)\sim d(E)e^{-\beta E}$, with $d(E)$ the density of states, but one must be careful in identifying $L$ directly with the temperature.
Instead, due to the generic exponential growth of the density of states in string theory, $d(E)\sim e^{\beta_H E}$, we need to identify $1/L=l_s(\beta-\beta_H)$, where $\beta_H \sim 1/M_s$ is the inverse Hagedorn temperature.
It follows that, upon injection of energy, the gas of highly excited strings mantains a temperature close to (and smaller than) the Hagedorn temperature, and this energy is used to populate the large number of highly excited states instead: we find a gas of long, highly excited strings at a temperature $T<T_H=1/\beta_H$.

\subsection{Moving out of equilibrium: equilibration rates}\label{sec:perturbations}

Boltzmann equations allow us to 
study thermodynamics out of equilibrium, which has important applications in cosmology.
In this section, we illustrate some non-equilibrium aspects of our toy model.
We will solve for the evolution of fluctuations $\delta n (x,t)$ in the ensemble, finding the rate at which these perturbations decay. 
The lessons learned from this toy model allow to compute equilibration rates of any system described by a Boltzmann equation.

Let us thus study the integro-differential equation obtained by linearizing~\eqref{eq:lowe-thorlacius} around the equilibrium solution~\eqref{eq:eq-dist-closed}, which we denote $\Bar{n}(x)$.
Namely, for $n(x,t)=\Bar{n}(x)+\delta n(x,t)$, we find at order $\mathcal{O}(\delta n)$ and letting the perturbation carry no energy:
\begin{equation}\label{eq:diffeq-f}
    \frac{1}{\kappa}\frac{\partial{ \delta n(x,t)}}{\partial t}=-\lr{\frac{x^2}{2}+xL}\delta n(x,t)+\int_0^x{dy\, y\delta n(y,t)\lr{e^{\frac{-(x-y)}{L}}-1}}\, .
\end{equation}

This equation admits a family of analytic solutions~\cite{Frey:2023khe}:
\begin{equation}\label{eq:zero-e-fluctuation}
    \delta n(x,t) = \sqrt{\frac{\kappa \pi(c+t L^2)}{2}}\frac{e^{-\frac{x}{L}+A(t)^2}}{L^2}\text{Erf}\lr{A(t),A(t)+\sqrt{\frac{\kappa (c+t L^2)}{2}}\frac{x}{L}}-\frac{e^{-x/L}}{L^2} \, ,
\end{equation}
where $c$ is an arbitrary constant, $\text{Erf}(z_1,z_2)=\frac{2}{\sqrt{\pi}}\int_{z_1}^{z_2}{e^{-t^2}dt}$ is the incomplete error function, and
\begin{equation}
    A(t)=\sqrt{\frac{\kappa (c+t L^2)}{2}}\lr{1-\frac{1}{\kappa (c+tL^2)}}\, ,
\end{equation}
up to an overall multiplicative constant that must be small to ensure $\delta n(x,t)\ll \Bar{n}(x)$.
Taking the limit $t\to \infty$, these solutions describe an exponentially decaying fluctuation with a length-dependent decay rate given as $\Gamma (x)=\kappa \lr{\frac{x^2}{2}+x L}$.

Being able to obtain an analytic description of the evolution of fluctuations is a feature of the zero-dimensional case which does not carry through to more complicated situations.
However, the main features of Eq.~\eqref{eq:zero-e-fluctuation} can be argued from simple considerations that can be applied to more realistic setups.
The starting point is Eq~\eqref{eq:diffeq-f}, which admits a simple physical interpretation: the first term encodes the absorption of the fluctuation by the bath (and so encodes the timescale of decay), while the second term describes the propagation of the fluctuation in length space.
We can thus read off the equilibration rate of the system described by $\bar{n}(x)$ from the factors that multiply its fluctuation $\delta n(x,t)$ at first order, for any system for which there is a Boltzmann equation description.

\section{The Hagedorn phase and its gravitational wave remnant}
\label{sec:thcosmo}
We now turn to the study of the Hagedorn phase in cosmology.
The setting is a universe with stringy energy density $\rho > M_s^4$ but with Hubble scale $H$ satisfying $H/M_s <1$ for the effective description of the evolution of the background to be under control
\footnote{String theory recovers the Einstein-Hilbert action of General Relativity as a first-order term in a higher-derivative expansion controlled in powers of $1/M_s$.}.
We will see that this condition can be satisfied, and that the energy density is distributed among highly excited open string degrees of freedom which radiate gravitons (and other massless closed strings) in an out-of-equilibrium way.
After dillution, the energy density is distributed into massless open string degrees of freedom, which would contain the SM in a concrete construction.
The Hagedorn phase therefore provides a natural reheating, together with predictions for a GW background.

In order to compute the GW spectrum, we need to understand the distribution of the strings sourcing GWs.
The kinetic theory developed in section~\ref{sec:stringthermo} is suited for this.
An important point is that string compactifications leading to a realistic phenomenology feature several sources of energy which backreact on the underlying geometry, rendering (strongly or not) warped metrics of the form $ds^2 =e^{2A(y)} g_{\mu\nu}dx^\mu dx^\nu+e^{-2A(y)}g_{mn}dy^mdy^n$ for some function $A(y)$, where $x$ are $3+1$ dimensional coordinates and $y$ are coordinates along the extra dimensions.
This effect localises the long strings in a small region in the extra dimensions, rendering the thermodynamics as effectively $3+1$ dimensional~\cite{Frey:2024jqy}.
Thus, the thermodynamic description of the system is largely independent of the physics in the rest of the extra dimensions - it is therefore generic.
Without entering into the details, a Boltzmann equation analysis as in section~\ref{sec:stringthermo} shows that:
\begin{itemize}
\item The energy density is dominated by long open string degrees of freedom, which source the expansion of the universe.
The important condition $M_s/H \gg 1$
\footnote{A similar parametric behaviour can accommodate the more stringent consistency condition $M_{KK}/H \gg 1$, with $M_{KK}$ the Kaluza-Klein scale defined by the size of the extra dimensions.},
can be satisfied provided there is a large hierarchy between the string and Planck scales, $M_s/M_p\ll 1$, which we assume in the following.
It is worth noting that this is the case in the most widely studied scenarios in string phenomenology~\cite{Kachru:2003aw,Balasubramanian:2005zx}.

\item The equilibration rates in the gas of open strings satisfy $\Gamma_{eq}/H \sim \sqrt{V}\gg 1$, where $V$ is the volume of the extra dimensions in string units.
Control over the low-energy, field theoretical description of string theory requires $V\gg 1$, so a consistent description in terms of Boltzmann equations implies that the long, open strings always equilibrate.

\item Gravitons do not equilibrate.
Graviton production rate is Planck-supressed, so the energy density in gravitons is never sufficient to reach thermal equilibrium.
Gravitons are therefore produced as an out-of-equilibrium process, as in the field theory setting of section~\ref{sec:SM}.
\end{itemize}

\subsection{Graviton emission and the GW spectrum}
We now turn to the emission rate of gravitons from typical strings and the GW spectrum from the Hagedorn phase.
The computation of graviton emission by typical strings in flat backgrounds has been carried out in the bosonic~\cite{Amati:1999fv} and supersymmetric~\cite{Kawamoto:2013fza} case at leading order in string perturbation theory in flat backgrounds
\footnote{
A technical subtlety regarding this computation has recently been raised~\cite{Firrotta:2024fvi}.
We showed~\cite{Frey:2024jqy} that the disagreement does not modify qualitatively the peak position or amplitude of the GW spectrum.}.
The general result is that the emission spectrum features a greybody spectrum at the Hagedorn temperature, $T_H \sim M_s$, suppressed by $(M_s/M_p)^2$.
Furthermore, \textit{this result is largely independent of the details of the compactification}~\cite{Frey:2024jqy}, rendering our results robust against model dependence.
The underlying reason is an averaging process required for the description of the thermodynamics.
The result for the production rate of gravitons  with frequencies between $\omega$ and $\omega +d\omega$ from a typical highly excited string of length $l$ is:
\begin{equation}\label{eq:grav-em}
\frac{d \Gamma_{o\to g}}{d\omega}= A \lr{\frac{M_s}{M_p}}^2 lM_s (\omega/T_H)^3 \frac{ e^{-\omega/T_H}}{\lr{1-e^{-\omega/2T_H}}^2}\, ,
\end{equation}
with $A$ a computable order one number.

With this production rate, we can write and solve an evolution equation for the energy density in gravitons to find the energy density today.
Eventually, neglecting model-dependent constants and assuming standard cosmology after the phase, the result for the GW spectrum is
\begin{eqnarray}
\label{eq:fidutial-gw-spectrum}
h^2\Omega_{\tti{GW},0} \simeq 10^{-10}
\lr{\frac{M_s}{10^{15} \,\text{GeV}}}
\lr{\frac{\omega_0}{100 \text{GHz}}}^{5/2} 
I\left(\frac{\omega_0}{100 \text{GHz}}\right)\, .
\end{eqnarray}

\begin{figure}[t!]
\centering
    \includegraphics[width=0.6\textwidth]{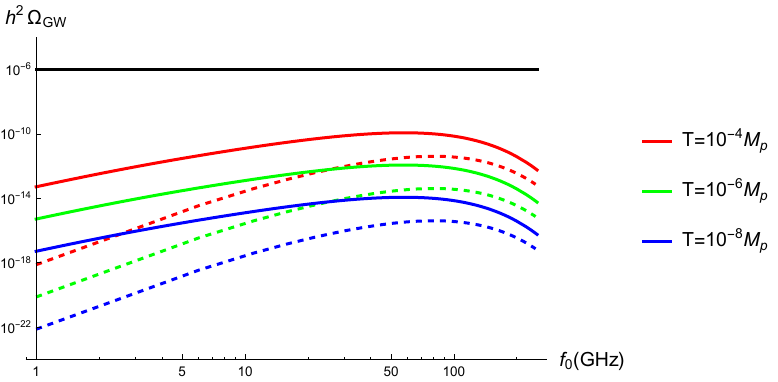}
    \caption{GW background sourced by the Hagedorn phase (solid) and the CGMB (dashed).
    At the same reheating temperature $T$ (taken to be the Hagedorn temperature), the Hagedorn phase predicts a larger amplitude and a smaller spectral index.}
    \label{fig:HAGEDORN}
\end{figure}

The spectrum is plotted in Fig.~\ref{fig:HAGEDORN}.
The position of the peak \textit{does not} depend on the local string scale and is determined by the cosmology following the Hagedorn phase -- for standard cosmological evolution after the epoch the peak is approximately 60 GHz.
 On the other hand, the overall strength of the amplitude is set by the
 local string scale and depends linearly on the ratio ${M_{s} \big{/} M_{p}}$. 

The behaviour of the peak frequency and amplitude resemble that of the SM as discussed in section~\ref{sec:SM}.
Indeed, one may think of both spectra as having the same origin: the GWs arising from a thermal phase in the early universe.
Our computation thus fixes the high-energy, stringy part of the spectrum and its amplitude turns out to be larger.
The reason is that the leading-order string process occurs at 3-points (is a decay), which, when allowed, is subdominant in field theory
\footnote{The process is suppressed by $m/M_p$ for a particle of mass $m$ decaying.
If $m \leq T$, the effect is negligible compared to $T/M_p$, and otherwise the number of particles decaying is exponentially suppressed.
These processes are thus subdominant in field theory.}.
The leading contribution in field theory involves four external legs instead and is therefore suppressed by higher powers of couplings.
The Hagedorn phase allows for efficient 3-point interactions due to the stringy feature of an exponentially growing density of states (which allows for states with masses larger than the temperature to be excited).

\section{Conclusions}
In this document, we have identified two sources of high-frequency gravitational waves in the early universe: the CGMB and the GW background from the Hagedorn phase.
    The first is field theoretical and UV sensitive, although its main features (order of magnitude of peak frequency and peak amplitude) are not modified qualitatively in extensions of the SM.
    The latter is stringy and requires a proper description of the Hagedorn phase, involving an ensemble of highly excited strings interacting with each other.
 We have described the kinetic theory of strings by means of Boltzmann equations~\cite{Frey:2023khe}.
This description has allowed us to accurately describe a Hagedorn phase in cosmology and compute its GW spectrum~\cite{Frey:2024jqy}, reaching the remarkable conclusion that the stringy prediction is largely independent of the details of the compactification and dominates over the field theory analogue.
    This is one of the few examples of a stringy signal dominating over the field theory prediction and illustrates the potential of GW astronomy as a way of testing high energy physics. 

\section*{Acknowledgments}

I would like to thank Andrew Frey, Ratul Mahanta, Anshuman Maharana, Francesco Muia and Fernando Quevedo for an exciting collaboration in these projects.
I also would like to extend my gratitude to Pham Quang Hung and the Rencontres du Vietnam for hospitality, in particular to The Nguyen and Thao Do, and the organizers of the 29$^{th}$ PASCOS conference for a wonderful organization.
This work was partially supported by STFC consolidated grant ST/T000694/1 and ST/X000664/1.
This article is based upon work from COST Action COSMIC WISPers CA21106, supported by COST (European Cooperation in Science and Technology).
\section*{References}

\bibliography{biblio}

\end{document}